\documentclass[
 preprint,
 superscriptaddress,
 amsmath,amssymb,
 aps,
 prl,
 longbibliography,
 showpacs
]{revtex4-1}
\usepackage{graphicx}
\usepackage{dcolumn}
\usepackage{bm}
\usepackage{epstopdf}
\begin{document}

\title{Low-threshold topological nanolasers based on second-order corner state}
\author{Weixuan Zhang}
\thanks{Contributed equally to this work.}
\affiliation{Key Laboratory of advanced optoelectronic quantum architecture and measurements of Ministry of Education, School of Physics, Beijing Institute of Technology, 100081, Beijing, China}
\affiliation{Beijing Key Laboratory of Nanophotonics $\&$ Ultrafine Optoelectronic Systems, Micro-nano Center, School of Physics, Beijing Institute of Technology, 100081, Beijing, China}
\author{Xin Xie}
\thanks{Contributed equally to this work.}
\affiliation{Beijing National Laboratory for Condensed Matter Physics, Institute of Physics, Chinese Academy of Sciences, Beijing 100190, China}
\affiliation{CAS Center for Excellence in Topological Quantum Computation and School of Physical Sciences, University of Chinese Academy of Sciences, Beijing 100049, China}
\author{Huiming Hao}
\affiliation{State Key Laboratory of Superlattices and Microstructures, Institute of Semiconductors Chinese Academy of Sciences, Beijing 100083, China}
\author{Jianchen Dang}
\author{Shan Xiao}
\author{Shushu Shi}
\affiliation{Beijing National Laboratory for Condensed Matter Physics, Institute of Physics, Chinese Academy of Sciences, Beijing 100190, China}
\affiliation{CAS Center for Excellence in Topological Quantum Computation and School of Physical Sciences, University of Chinese Academy of Sciences, Beijing 100049, China}
\author{Haiqiao Ni}
\affiliation{State Key Laboratory of Superlattices and Microstructures, Institute of Semiconductors Chinese Academy of Sciences, Beijing 100083, China}
\author{Zhichuan Niu}
\email{zcniu@semi.ac.cn}
\affiliation{State Key Laboratory of Superlattices and Microstructures, Institute of Semiconductors Chinese Academy of Sciences, Beijing 100083, China}
\author{Can Wang}
\author{Kuijuan Jin}
\affiliation{Beijing National Laboratory for Condensed Matter Physics, Institute of Physics, Chinese Academy of Sciences, Beijing 100190, China}
\affiliation{CAS Center for Excellence in Topological Quantum Computation and School of Physical Sciences, University of Chinese Academy of Sciences, Beijing 100049, China}
\affiliation{Songshan Lake Materials Laboratory, Dongguan, Guangdong 523808, China}
\author{Xiangdong Zhang}
\email{zhangxd@bit.edu.cn}
\affiliation{Key Laboratory of advanced optoelectronic quantum architecture and measurements of Ministry of Education, School of Physics, Beijing Institute of Technology, 100081, Beijing, China}
\affiliation{Beijing Key Laboratory of Nanophotonics $\&$ Ultrafine Optoelectronic Systems, Micro-nano Center, School of Physics, Beijing Institute of Technology, 100081, Beijing, China}
\author{Xiulai Xu}
\email{xlxu@iphy.ac.cn}
\affiliation{Beijing National Laboratory for Condensed Matter Physics, Institute of Physics, Chinese Academy of Sciences, Beijing 100190, China}
\affiliation{CAS Center for Excellence in Topological Quantum Computation and School of Physical Sciences, University of Chinese Academy of Sciences, Beijing 100049, China}
\affiliation{Songshan Lake Materials Laboratory, Dongguan, Guangdong 523808, China}
\begin{abstract}

The topological lasers, which are immune to imperfections and disorders, have been recently demonstrated based on many kinds of robust edge states, being mostly at microscale. The realization of 2D on-chip topological nanolasers, having the small footprint, low threshold and high energy efficiency, is still to be explored. Here, we report on the first experimental demonstration of the topological nanolaser with high performance in 2D photonic crystal slab. Based on the generalized 2D Su-Schrieffer-Heeger model, a topological nanocavity is formed with the help of the Wannier-type 0D corner state. Laser behaviors with low threshold about 1 $\mu W$ and high spontaneous emission coupling factor of 0.25 are observed with quantum dots as the active material. Such performance is much better than that of topological edge lasers and comparable to conventional photonic crystal nanolasers. Our experimental demonstration of the low-threshold topological nanolaser will be of great significance to the development of topological nanophotonic circuitry for manipulation of photons in classical and quantum regimes.

\end{abstract}
\maketitle


The investigation of topological photonics has become one of the most fascinating frontiers in recent years \cite{haldane2008possible,lu2014topological,ozawa2019topological,wang2009observation,
fang2012realizing,rechtsman2013photonic,khanikaev2013photonic,hafezi2013imaging,chen2014experimental,
wu2015scheme,barik2018topological,mittal2018topological,tambasco2018quantum,wang2019direct}. In addition to the conventional passive and linear system, exploring the topological phenomena in highly nonlinear environments also possesses significant influences \cite{bahari2017nonreciprocal,harari2018topological,bandres2018topological,zhao2018topological,parto2018edge,st2017lasing,
ota2018topological,shao2020high,Zeng2020Electrically,smirnova2019nonlinear}. Recently, the concept of topological lasers, whose lasing mode exhibits topologically protected transportation with the help of robust 1D edge state in 2D systems, is proposed and demonstrated \cite{bahari2017nonreciprocal,harari2018topological,bandres2018topological}. The pioneer work is the study of lasing on the nonreciprocal topological cavities, which are formed by a closed quantum-Hall liked edge state, at telecommunication wavelengths \cite{bahari2017nonreciprocal}. While, due to the weak magneto-optic effect, the topological band-gap is only about 40 $pm$. The first magnet-free scheme for the realization of single-mode topological lasers is based on an array of ring resonators in 2D, where the notably higher slope efficiencies is observed compared to the trivial counterparts \cite{harari2018topological,bandres2018topological}. Besides the utilization of 1D edge state, the 0D boundary states existing in 1D lattices with non-trivial Zak phases \cite{zhao2018topological,parto2018edge,st2017lasing,ota2018topological} and the topological bulk state around the band edge \cite{shao2020high} have also been used to realize topological lasers. However, the currently designed topological lasing systems are almost at microscale, so that the corresponding thresholds are usually about several milliwatts. The topological nanolasers, combining the advantages of topological robustness and nanolasers including the small footprint, low threshold and high energy efficiency \cite{ota2017thresholdless,jang2015sub,takiguchi2016systematic,strauf2011single,painter1999two,park2004electrically,hamel2015spontaneous,
noda2001polarization,yu2017demonstration,yoshida2019double}, are still lacking except for the scheme using 0D interface state in the 1D photonic beam with threshold being about 46 $\mu W$ \cite{ota2018topological}.

Recently, a new class of symmetry-protected higher-order topological insulators, which sustain lower-dimensional boundary states and obey a generalization of the standard bulk-boundary correspondence, have been proposed \cite{benalcazar2017quantized,imhof2018topolectrical,peterson2018quantized,serra2018observation,mittal2019photonic,
langbehn2017reflection,ezawa2018higher,schindler2018higher,xie2019visualization,chen2019direct,ota2019photonic,
xue2019acoustic,zhang2019second,noh2018topological,liu2019second,ni2019observation}. In 2D cases, the 0D corner state can be usually formed by two mechanisms. One is related to quantized bulk quadrupole polarization \cite{benalcazar2017quantized,imhof2018topolectrical,peterson2018quantized,serra2018observation,mittal2019photonic} and another is derived from the edge dipole polarization quantized by the 2D Zak phase \cite{xie2019visualization,chen2019direct,ota2019photonic}. The latter model can be easily implemented in the compact magnet-free optical platform \cite{chen2019direct} and be used to construct topological nanocavity \cite{ota2019photonic}. The problem is whether we can exploit the topological nanocavity, sustaining the high quality (Q) factor and small mode volume comparable to the conventional photonic crystal nanocavity \cite{akahane2003high}, to realize the topological nanolaser with low threshold and high energy efficiency.

In this work, we report on the first experimental demonstration of the topological nanolaser in 2D topological photonic crystal (PhC) nanocavity, which sustains the Wannier-type 0D corner state at nanoscale. Our designed topological nanocavity is based on the 2D Su-Schrieffer-Heeger (SSH) model, in which the corner state is induced by the edge dipole polarization quantized by the 2D Zak phase. By suitably tuning the gap distance between the trivial and non-trivial parts of the PhC slab, the higher Q factor can be achieved. The robustness of corner state with respect to defects in the bulk of PhC is demonstrated. Lasing behaviors at corner state with high performance including low threshold and high spontaneous emission coupling factor ($\beta$), are observed with InGaAs quantum dots (QDs) serving as active material. The high performance of the topological nanolaser is comparable to the conventional semiconductor nanolasers \cite{ota2017thresholdless,jang2015sub,takiguchi2016systematic,strauf2011single}, indicating the great prospects of the topological nanocavity for a wide range of applications in the topological nanophotonic circuitry.


\begin{figure}
\centering
\includegraphics[scale=0.63]{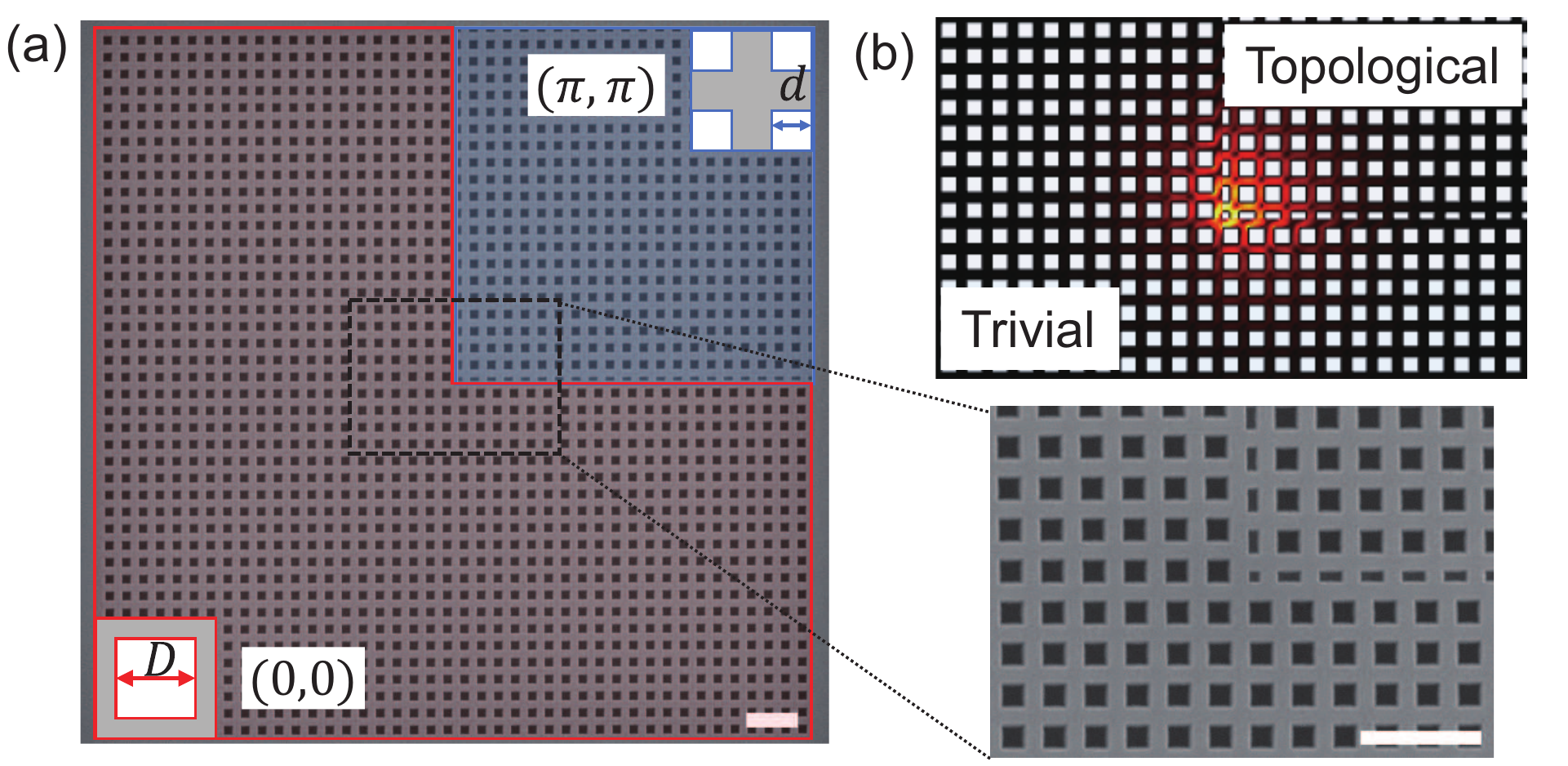}
\caption{(a) Scanning electron microscope image of a fabricated 2D topological PhC cavity in a square shape. Inset on the right shows an enlarged image around the corner. The scale bar is 1 $\mu m$. The topological nanocavity consists of two topologically distinct PhC which are indicated by the red and blue areas. They have different unit cells as shown in insets. $d$ and $D$ are the lengths of the squares in blue and red unit cells, in which $D=2d$. (b) Electric field profile of topological corner state calculated by the finite element method \cite{strang1973analysis}. }
\label{f1}
\end{figure}

Based on the generalized 2D SSH model, a topological nanocavity can be constructed, as shown in Fig. \ref{f1}(a). It consists of two kinds of PhC in a square shape, which have the same period $a$ but different unit cells, as indicated by the red and blue areas in Fig. \ref{f1}(a). These two regions share the common band structure but possess different topologies, which are characterized by the 2D Zak phase $\theta^{Zak}$, a quantity defined by the integration of Berry connection within the first Brillouin zone \cite{zak1989berry,liu2017novel}. The PhC in blue (red) area has a nontrivial (trivial) 2D Zak phase of $\theta^{Zak}=(\pi,\pi)$ ($\theta^{Zak}=(0,0)$). In this case, according to the bulk-edge-corner correspondence, the midgap 0D corner state can be induced at the intersection of two boundaries with non-zero edge polarizations, which are protected by the nontrivial 2D Zak phase in the bulk. Furthermore, the eigen-field of this corner state is highly confined at nanoscale as show in Fig. \ref{f1}(b), which can greatly enhance the light-matter interaction, thus having potential applications such as the construction of topological nanolasers.

\begin{figure}
\centering
\includegraphics[scale=0.63]{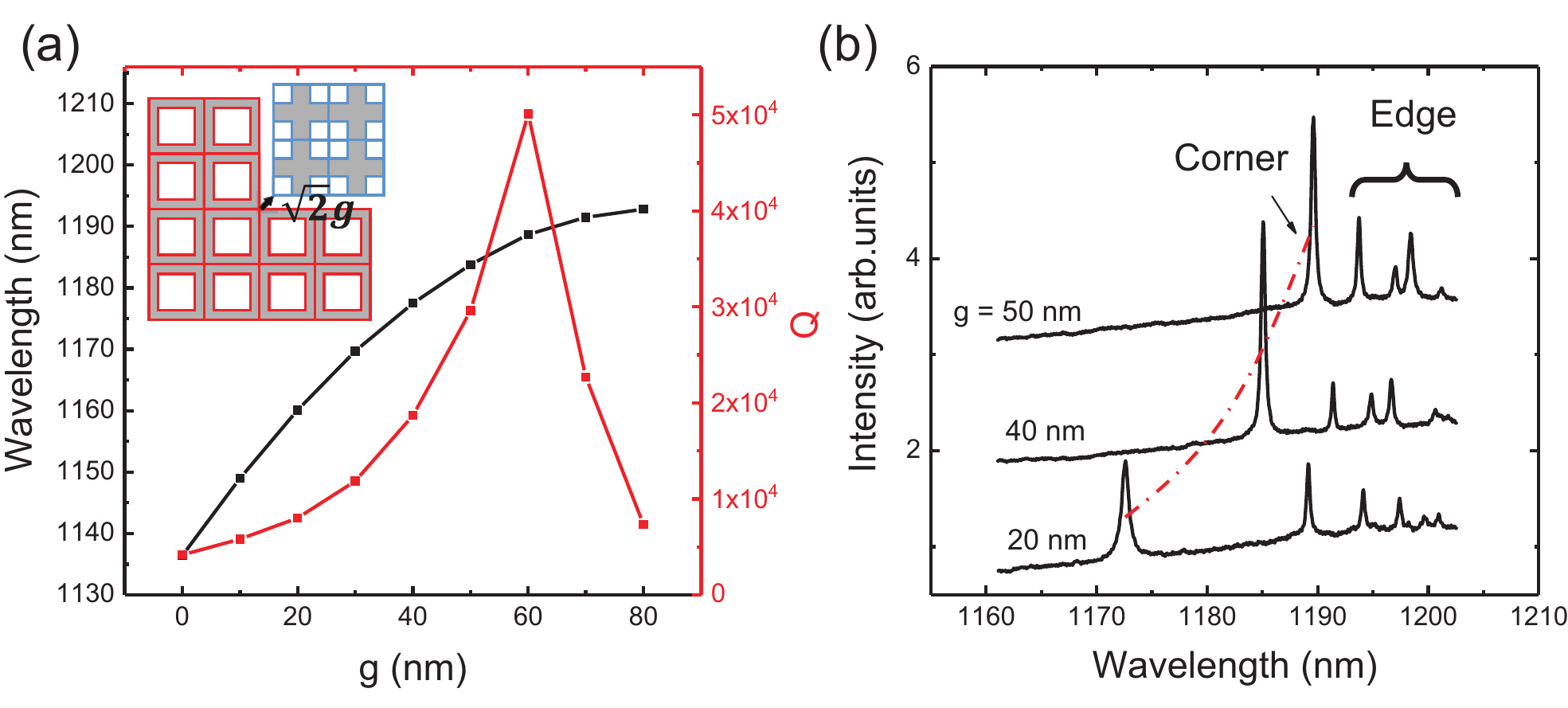}
\caption{ (a) Calculated Q factors (red) and wavelengths (black) of corner state with different $g$. The Q factors and wavelengths are calculated by the finite element method. Other parameters for these cavities are that $a=380$ nm, $D=242$ nm. Inset shows the schematic of Q optimization, in which the topological PhC is shifted away from the corner by $\sqrt{2}g$ along the diagonal direction. (b) PL spectra for cavities with $a=380$ nm, $D=242$ nm and different $g$. Red dashed line represents the corner state. These peaks in long-wavelength range originate from edge states. The PL spectra are shifted for clarity.}
\label{f2}
\end{figure}

To improve the laser performance, the Q factor of corner state is optimized by suitably tuning the gap distance ($g$) between the trivial and non-trivial parts of the PhC slab as shown in the inset of Fig. \ref{f2}(a). The black and red lines in Fig. \ref{f2}(a) represent the calculated results of Q factor and resonance wavelength of the corner state with different values of $g$. It is clearly shown that the Q factor first increases then decreases with $g$ gradually going up, meanwhile, the corner state shows a redshift. When $g=60$ nm, the corner mode supports a high Q factor of 50,000 and a small mode volume of $0.61(\lambda/n)^3$. It is worthy to note that the Q factor and mode volume of corner state are both able to be disturbed by introducing perturbations around the corner. Nevertheless, the corner state always survives even with harsh perturbations on the bulk of PhC as far as the topological properties of the bulk band are not changed, which could be a practical advantage for robust applications.


We fabricated the designed topological nanocavity with different parameters into a 160-nm-thick GaAs slab using electron beam lithography followed by inductively coupled plasma and wet etching process. The wet etching with HF solution was used to remove the sacrificial layer to form air bridge. The GaAs slab is grown by molecular beam epitaxy and contains a single layer of InGaAs QDs at the center with a density of about 500 $\mu m^{-2}$. The scanning electron microscope image of a fabricated cavity is shown in Fig. \ref{f1}(a), in which the inset on the right shows an enlarged image around the corner.

To optically characterize the corner state, we performed the confocal micro-photoluminescence (PL) measurements at 4.2 K using a liquid helium flow cryostat. An objective lens with a numerical aperture of 0.7 was used to address the sample. It was excited by a continuous laser with wavelength of 532 nm. The PL signal was dispersed by a grating spectrometer and detected with a liquid-nitrogen-cooled charge coupled device camera. Fig. \ref{f2}(b) shows PL spectra for cavities with different $g$, in which the corner states and edge states are indicated. With increasing $g$ from 20 nm to 50 nm, the corner state exhibits redshift and the corresponding Q factor increases, which are consistent with the numerical results in Fig. \ref{f2}(a). However, limited by the unavoidable fabrication imperfections, the fabricated Q factors are about an order of magnitude lower than the theoretical prediction, which are usually 2500-5000.

\begin{figure}
\centering
\includegraphics[scale=0.63]{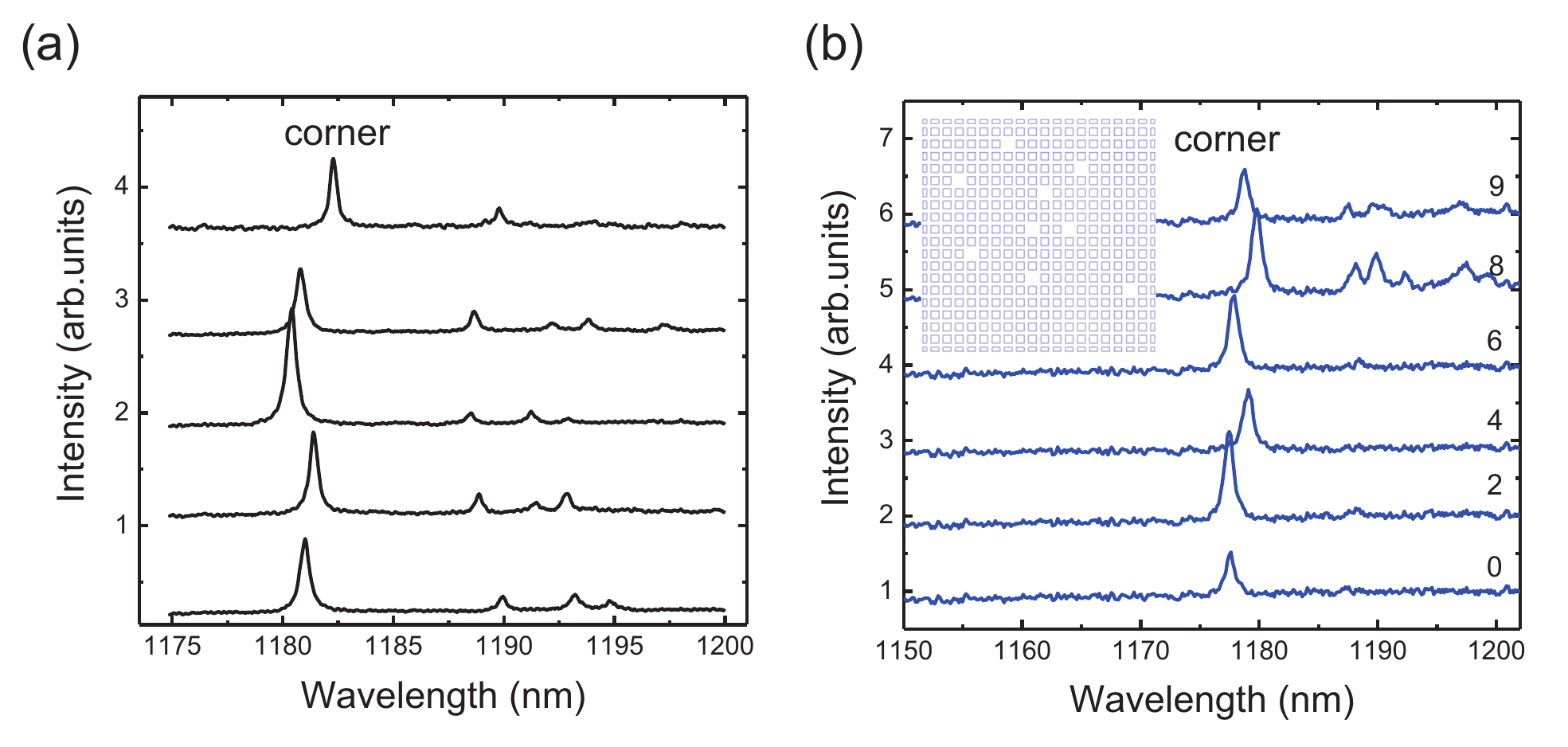}
\caption{ (a) PL spectra of defect-free cavities with the parameters of $a=380$ nm, $D=242$ nm and $g=50$ nm. (b) PL spectra of cavities with different numbers of defects as shown in inset. The numbers represent the number of missing square holes in bulk of PhC. Here, the missing square holes are several periods away from the corner. The PL spectra are shifted for clarity.}
\label{f3}
\end{figure}

Then, in order to demonstrate the topological protection of the corner state, we fabricated topological cavities without and with defects introduced by missing square holes in bulk of PhC (Inset in Fig. \ref{f3}(b)). Fig. \ref{f3}(a) shows the PL spectra for defect-free cavities. In this case, the fluctuation of corner state and edge state is observed, which may result from the fabrication imperfections. In the current state-of-the-art techniques, a fabrication imperfection of about 2-5 nm always exists. According to the calculated results (see Supplementary Material), the fluctuation of resonance wavelength can be up to 6 nm when random perturbation about 2-5 nm are introduced around the corner. The observed fluctuation of wavelength about 2 nm is within the range. Additionally, the calculated Q factor with perturbations around corner can be decreased by about an order of magnitude, and that can well explain the deviation of fabricated Q factors from the calculated result.

The corner state is induced by quantized edge polarizations, which is topologically protected by the nontrivial 2D Zak phases of bulk band. Therefore, the corner state exists even with harsh perturbations as long as the topological property of PhC is not changed. It is worthy to note that the Q and wavelength are not topologically protected in theory. To demonstrate the topological robustness of corner state, we introduced defects in bulk of PhC by missing square holes, as shown in the inset of Fig. \ref{f3}(b). PL spectra of cavities with different amounts of defects in the bulk of PhC are shown in Fig. \ref{f3}(b). The corner state still exists even with nine missing holes except for small fluctuations of Q factors and wavelengths. The wavelength fluctuation is about 2.5 nm, which is comparable to the defect-free samples in Fig. \ref{f3}(a). According to the numerical results, the wavelength deviation of cavities with 9 missing holes is about 0.8 nm, which can be ignored in comparison with the deviation induced by fabrication imperfection. The fluctuation in cavities with defects thus mainly results from the fabrication imperfections. Therefore, the existence of corner state is demonstrated to be robust to the defects in the bulk of PhC.

\begin{figure}
\centering
\includegraphics[scale=0.63]{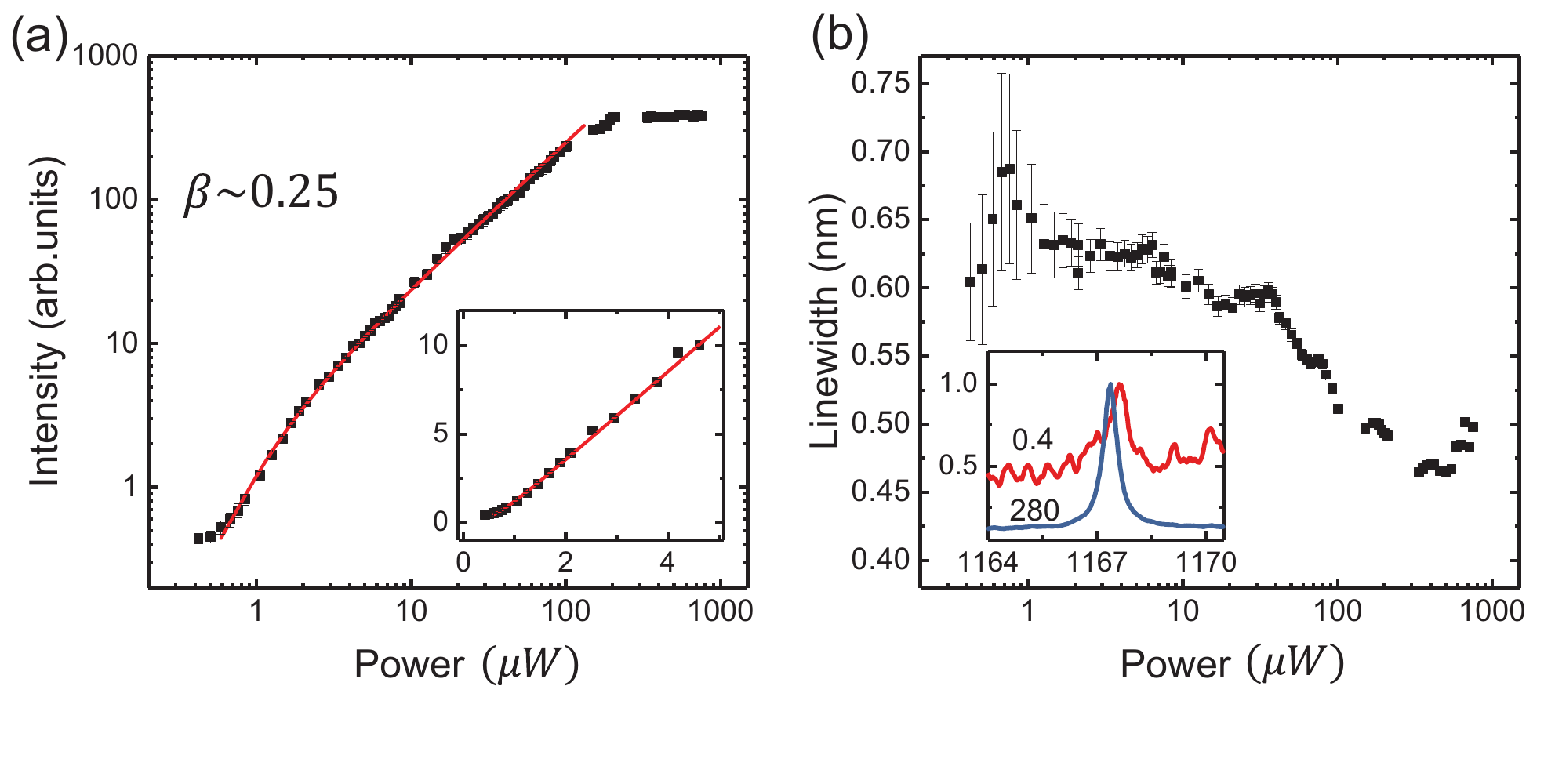}
\caption{(a) Pump-power dependence of corner state from cavity with $a=360$ nm, $D=222$ nm and $g=30$ nm, which are on a logarithmic scale. Insets show the enlarged curve around threshold on a linear scale. Squares represent the experimental data and the line represents the fitted result by semiconductor laser model. $\beta$ for Cavity a is estimated as about 0.25. Their lasing thresholds are about 1 $\mu W$. (b) Linewidths of corner state as function of pump power. Insets show the normalized PL spectra with different pump powers. The unit of pump powers is $\mu W$. Linewidth shows a clear narrowing. The linewidths and intensities are both extracted by fitting high-resolution spectra with Lorentz peak functions. }
\label{f4}
\end{figure}

To verify the laser behavior of corner state with QDs as gain, we investigate the pump-power dependence of corner state emission in the topological nanocavities. In this case, the continuous laser with wavelength of 532 nm was used to pump QDs. We have measured many cavities and observed the lasing and non-lasing behaviors (see Supplementary Material). Fig. \ref{f4} illustrates lasing behaviors with high performances from cavity with $a=360$ nm, $D=222$ nm and $g=30$ nm. The linewidths and intensities of corner state are both extracted by fitting high-resolution spectra with Lorentz peak functions. The light-in-light-out (L-L) plot on logarithmic scale in Fig. \ref{f4}(a) shows a mild `s' shape, suggesting a lasing oscillation with high $\beta$. A clear kink is observed in the L-L curve on a linear scale, indicating a low threshold about 1 $\mu W$. The $\beta$ factor is extracted by fitting the curve with semiconductor laser model \cite{bjork1991analysis} (see Supplementary Material), which is about 0.25. The observed thresholds of our proposed higher-order topological nanolaser are about three orders of magnitude lower than that of the current topological edge lasers \cite{bandres2018topological,bahari2017nonreciprocal,zhao2018topological,parto2018edge,st2017lasing}, furthermore are around two percent of the threshold in the topological nanolaser based on 0D interface state \cite{ota2018topological}.

In addition, the measured linewidths in Fig. \ref{f4}(b) exhibit a clear decline and spectral narrowing in PL are shown in insets, further verifying the laser behaviors in the topological nanocavities. The pronounced linewidth re-broadening around the threshold is observed, which is a signature for lasing in PhC QD nanolaser \cite{strauf2011single}. It is worthy to note that the saturation of intensities and increase of linewidths at high pump power may result from heating in nanocavities. At pump power about 0.5 $\mu W$ which is below the thresholds, the Q factor of the cavity mode is estimated as about 1700. Therefore, the low threshold and high $\beta$ can be attributed to the small volume and high Q factor which lead to strong optical confinement. The high performance of the topological nanolaser is comparable to that of conventional nanolasers \cite{ota2017thresholdless,jang2015sub,takiguchi2016systematic,strauf2011single}, indicating the great prospect in applications with built-in protection.


In summary, we demonstrated the topological nanolaser with high performance based on the second-order topological corner state in 2D PhC slabs for the first time. The Q factor of corner state has been optimized by suitably tuning the distance between topologically distinct PhC slabs, which is confirmed both theoretically and experimentally. The topological protection of corner state is demonstrated by introducing defects in the bulk of PhC. The laser behavior with low threshold about 1 $\mu W$ and high $\beta$ about 0.25, is observed. The observed thresholds are much lower than current topological lasers due to small mode volume and high Q factor, and the performance is comparable to the conventional nanolasers. Our result shows an example on downscaling the applications of topological photonics into nanoscale and demonstrates the great potential of the topological nanocavity on the applications in topological nanophotonic devices.

\begin{acknowledgments}
This work was supported by the National Natural Science Foundation of China (Grants No. 11934019, No.11721404, No. 51761145104, No. 61675228, and No. 11874419), the National key R$\&$D Program of China (Grant No. 2017YFA0303800 and No. 2018YFA0306101), the Key R$\&$D Program of Guangdong Province (Grant No. 2018B030329001), the Strategic Priority Research Program (Grant No. XDB28000000), the Instrument Developing Project (Grant No. YJKYYQ20180036) and the Interdisciplinary Innovation Team of the Chinese Academy of Sciences.
\end{acknowledgments}

\end{document}